\documentclass[twocolumn,twocolappendix,tighten]{aastex63}

\usepackage{amssymb, amsmath}
\usepackage{comment}
\usepackage{hyperref}
\usepackage{amssymb, amsmath}
\def\vb#1{\mbox{\boldmath $#1$}}

\usepackage{here}
\usepackage{ulem}

\shorttitle{Amplification and saturation of turbulent magnetic field}
\shortauthors{Higashi, Susa, Federrath, \& Chiaki}

\begin{document}
\title{Amplification and saturation of turbulent magnetic field in collapsing primordial gas clouds}

\author[0000-0003-1029-7592]{Sho Higashi}
\affiliation{Department of Physics, Konan University, Okamoto, Kobe, Japan}
\author[0000-0002-3380-5302]{Hajime Susa}
\affiliation{Department of Physics, Konan University, Okamoto, Kobe, Japan}
\author[0000-0002-0706-2306]{Christoph Federrath}
\affiliation{Research School of Astronomy and Astrophysics, Australian National University, Canberra, ACT~2611, Australia}
\affiliation{Australian Research Council Centre of Excellence in All Sky Astrophysics (ASTRO3D), Canberra, ACT~2611, Australia}
\author[0000-0001-6246-2866]{Gen Chiaki}
\affiliation{National Astronomical Observatory of Japan, Mitaka, Tokyo, Japan}
\affiliation{National Institute of Technology, Kochi College, Monobe, Kochi, Japan}


\begin{abstract}
Recent numerical studies suggest that magnetic fields play an important role in primordial star formation in the early universe.
However, the detailed evolution of the magnetic field in the collapse phase still has uncertainties because of the complicated physics associated with turbulence in a collapsing magnetized system.
Here, we perform a suite of numerical MHD simulations that follow the collapse of magnetized, turbulent primordial gas clouds to investigate the evolution of the magnetic field associated with the turbulence, assuming a polytropic equation of state with exponent $\gamma_{\rm eff}$ and with various numerical resolutions.
In addition, we generalize the analytic theory of magnetic field growth/saturation so that it can deal with various exponents $\gamma_{\rm eff}$ and turbulence energy spectra.
We find that the numerical results are well reproduced by the theory for various $\gamma_{\rm eff}$ through the collapse phase during the formation of the first stars.
The magnetic field is eventually amplified by a factor of $10^{12}$ -- $10^{15}$ due to kinematic and non-linear turbulent dynamo effects and reaches 3\% -- 100\% of the equipartition level, depending on $\gamma_{\rm eff}$.
We also find that the transition between the kinematic and non-linear stages can be analytically estimated. 
These results indicate that the strong magnetic field accompanied by supersonic turbulence is a general property and suggest that it can play a crucial role in the formation of the first stars.
\end{abstract}


\section{Introduction} \label{sec:intro}
Theoretical studies in the $\Lambda$CDM paradigm predict that the first generation of stars, known as population III stars, are formed in host dark matter minihalos at redshifts $z \gtrsim 10$ \citep{Haiman96,Tegmark97,Nishi99, Fuller00,Abel02,Bromm02,Yoshida03}.
In such a primordial environment, the star-forming gas only contains light elements.
The cooling rates per volume due to the chemical species of light elements (e.g., ${\rm H_2}$) are lower than those of heavier elements (e.g., CO).
As a result, stars are born in an environment 2--3 orders of magnitude higher in mass and 2 orders of magnitude higher in temperature than the present-day star-forming environment, such as the Milky Way.
Therefore, the stars are expected to be typically high mass $(\sim 10-1000 M_{\odot})$ due to the large fragmentation scale and high mass accretion rate if they form as a single stars in the cloud \citep[e.g.,][]{Susa96,Omukai98,Yoshida08}.

On the other hand, a number of recent numerical studies have indicated that the first stars are likely formed as multiple systems, such as a binary, triplet, etc.
The multiple systems form through the fragmentation of the accretion disk, which emerges after the formation of the central protostar 
\citep{Greif11,Greif2012,Susa13,Susa14,Susa19,Hirano17,Inoue20, Sharda20, Sharda21_dynamo, Chiaki22, Sugimura20, Sugimura23}.
Studying this fragmentation behavior is considered to be crucial for determining the initial mass function (IMF) of the first stars.  

Turbulence and magnetic fields are fundamental physical phenomena in the study of the disk fragmentation.
For instance, disk fragmentation is promoted by turbulence \citep{Clark11,Riaz18,Wollenberg20, Sugimura23} or suppressed by magnetic fields \citep{Machida13,Sharda20,Sadanari21}.
The efficiency of these effects depends on the properties of the magnetic field (its strength and structure); therefore, investigating and clarifying their details are essential for the study of the first-star formation.

As for the turbulence, previous studies indicated that it is amplified by gravitational contraction \citep[e.g.,][]{Vazques-semadeni98, Sur10, Sur12, Federrath11, Robertson2012, Birnboim18, Mandal20, Hennebelle21}.
In addition to these studies, we have recently confirmed that the turbulence becomes supersonic, even if the initial velocity fluctuations are relatively small at the onset of a collapse phase \citep{HSC21}.  
We also found that the amplified turbulent velocities are saturated with a constant Mach number by a balance between the energy injection via gravitational contraction and dissipation \citep{HSC22}.
Since the subsonic turbulence with a Mach number $\mathcal{M} \sim 0.5$ is often generated due to the accretion flow of gas along the filaments into halos at the onset of the cloud collapse \citep{Wise07, Greif08, Klessen10}, the gas cloud cores in minihalos may usually have saturated supersonic turbulent velocities during the collapse phase.

As for the magnetic field, one of the representative generation mechanisms of a very weak seed magnetic field in the early universe is introduced by \citet[so-called Biermann battery mechanism]{Biermann50}.
In the primordial environment, a non-parallel component of the gradient of electron density and pressure due to the virial shock at the halo formation or the spatial variations of ionization in the presence of temperature gradients can generate the seed field ($\sim 10^{-18}$~G) \citep[e.g.,][]{Kulsrud92, Lazarian92, Gnedin00},
or strong radiation flux from the neighboring star-forming clouds may also generate the seed field \citep{Ando10,Doi11}.
Although this magnetic field was considered to be not dynamically important for the first-star formation, recent studies with an analytical/numerical approach show that the turbulent motions of gas convert the kinetic energy into magnetic energy by randomly stretching, twisting, and folding the magnetic field lines on the viscous scale.
This is so-called the small-scale (turbulent) dynamo effect\citep[e.g.,][]{Brandenburg05, Schekochihin07, Federrath16}.
Due to the short time scales associated with the small spatial scales, the dynamo operates highly efficiently and increases the magnetic energy with an exponential growth rate \citep[e.g.,][]{Kazantsev68, Kulsrud92, Federrath11_turb, Schober15, Xu16}. 
In collapsing primordial gas clouds, earlier numerical studies show that the magnetic field is amplified due to both a small-scale dynamo and due to compression during the collapse \citep[e.g.,][]{Sur10,Sur12, Federrath11,Turk2012}.
Additionally, recent studies further investigate the detailed properties of the evolution of the magnetic field in collapsing gas clouds \citep{Xu20,Sharda21_dynamo, Hirano22,Mckee20, Stacy22, Sadanari23}.

In this paper we comprehensively investigate the growth/saturation of magnetic fields due to turbulence for various effective polytropic exponent $\gamma_{\rm eff}$ of the barotropic EoS, $P\propto \rho^{\gamma_{\rm eff}}$, by means of high-resolution numerical simulations and analytic calculations.
We show that the magnetic field always grows to an equipartition level in the collapsing primordial star-forming cloud.

This paper is structured as follows. We describe the setup of the numerical simulations in Section~\ref{sec:method} and present our results in Section~\ref{sec:results}. 
We introduce earlier analytical/theoretical estimates and generalize them in Section~\ref{sec:estimate}.
We compare the analytic calculations to our simulation results in Section~\ref{sec:comparison}, and present the results of the numerical resolution study in Section~\ref{sec:resolution}.
Section~\ref{sec:discussion} is devoted to discussion. We finally summarize the main points of this work in Section~\ref{sec:summary}.

\section{Numerical Method}\label{sec:method}

We perform a suite of high-resolution three-dimensional collapse simulations by using the adaptive mesh refinement (AMR) magneto-hydrodynamics (MHD) code \textsc{athena++} \citep{Athenapp}\footnote{\url{https://github.com/PrincetonUniversity/athena}}.
In order to solve the MHD equations with self-gravity in Cartesian coordinates, we choose a 2nd-order accurate Piecewise Linear Method (PLM) for the spatial reconstruction and a 2nd-order accurate Runge-Kutta method for the time integration, using an HLLD Riemann solver \citep{Miyoshi05}.
The basic equations are
\begin{eqnarray}
\frac{\partial \rho}{\partial t} + \nabla \cdot (\rho \vb{v}) &=& 0,  \label{eq:beq1}\\
\frac{\partial \rho \vb{v}}{\partial t} + \rho(\vb{v}\cdot \nabla)\vb{v} &+& \nabla P = \nonumber \\
&&(\vb{B}\cdot \nabla)\vb{B} -\rho\nabla\phi,  \label{eq:beq2}\\
\frac{\partial E}{\partial t} + \nabla \cdot [\left(E + P \right)\vb{v} &-&\vb{B}(\vb{B}\cdot \vb{v})]=\nonumber \\
&-&\rho \vb{v} \cdot \nabla \phi -\Lambda + \Gamma, \label{eq:beq3}\\
\frac{\partial \vb{B}}{\partial t} - \nabla \times (\vb{v}\times \vb{B}) &=& 0.
\label{eq:indeq}
\end{eqnarray}
In these equations, $E, ~ \rho, ~ \vb{v}, ~ \vb{B}$, and $\phi$ are the total energy density, mass density, velocity, magnetic field, and gravitational potential, respectively.
The total energy density is given by $E = e + \rho v^2 /2 + B^2/2$, where $e$ is the thermal energy density.
The total pressure $P$ is composed of $P = p + B^2/2$, where $p$ is the gas thermal pressure.
The gravitational potential $\phi$ is obtained from the Poisson's equation, using the full multigrid (FMG) method, implemented by \cite{Tomida23} with the multipole boundary condition,
\begin{equation}
    \nabla^2\phi = 4\pi G\rho.
\end{equation}
To satisfy the divergence-free constraint of the magnetic field $\left(\nabla \cdot \vb{B} = 0 \right)$, \textsc{athena++} adopts the constrained transport (CT) scheme.
It enables us to preserve the divergence of the magnetic field to zero, to the machine precision.

Our simulations in this paper are performed with the assumption of ideal MHD.
This means that we explicitly neglect the non-ideal MHD effects of the magnetic field, such as Ohmic dissipation and/or ambipolar and Hall diffusion, because these dissipation effects can be safely neglected, at least at the core scale during the collapse phase of the primordial gas cloud \citep{Maki&Susa2004, Maki07, Susa2015, nakauchi19,Sadanari23}.

We also assume a barotropic Equation of State (EoS) for simplicity.
In this model, the temperature evolution of the collapsing gas is calculated by using a polytropic model as in \cite{HSC21,HSC22}.
The relation between the thermal pressure $p$ and the density $\rho$ of the gas is expressed as
\begin{equation}
    p \propto \rho^{\gamma_{\rm eff}},
\end{equation}
with the effective polytropic exponent $\gamma_{\rm eff}$.
We perform simulations with 3~difference polytropic indices of
\begin{eqnarray}
\gamma_{\rm eff} = \begin{cases}
    1.2, &  \\
    1.1 & (\text{primordial}), \\
    1.0 & (\text{isothermal}).
  \end{cases}
\end{eqnarray}
Here, the model for $\gamma_{\rm eff}=1.1$ mimics the primordial gas in minihalos \citep{Omukai98}.
The other two models are used for comparison.

As an initial condition, we generate a gravitationally unstable, magnetized Bonner-Ebert sphere with a central density $\rho_{\rm peak,0}=4.65\times 10^{-20} ~\mathrm{g ~ cm^{-3}}$, and a uniform temperature of $T_{0}=200 ~ \mathrm{K}$.
The initial cloud mass is $1700 M_{\odot}$.
The computational domain has a size of 5 times the initial cloud radius with periodic boundary conditions.

By following earlier numerical results \citep[e.g.,][]{Federrath10, Federrath13, Federrath16_arxiv}, we add an initial velocity field with a turbulence spectrum typical of supersonic turbulence, $E(k) \propto k^{-2}$ \citep{BURGERS1948}.
Here, $E(k)$ is the 1D turbulent kinetic energy power spectrum with wavenumber $k$ in Fourier space.
This turbulent velocity field corresponds to a root-mean-square Mach number of 0.5, and is composed of a mixture of solenoidal (divergence-free) and compressive (solenoidal-free) modes, with a 2:1 ratio, called the natural mixture \citep{Federrath08, Federrath10}.

We also add a uniform magnetic field along the $z$-axis with the initial strength $\vb{B}_0=10^{-9}$~G.
Note that the strength and statistical properties of the dynamo-generated magnetic fields are independent of the seed field structure \citep{Seta20}. Additionally, theoretical studies \citep[e.g.,][]{Mckee20} have shown that the very weak seed field ($\lesssim 10^{-19}$ G) generated by the Biermann battery \citep{Biermann50} can be amplified quickly to moderate values ($\sim 10^{-8}$ G) due to the turbulent dynamo.
Hence our choice of initial magnetic field strength and structure is reasonable, but not critical to the final magnetic field strength and structure.

All of our numerical simulations start with a base grid of $512^3$ grid cells.
Using the AMR technique, we enforce to resolve the Jeans length by progressively refining cells as the collapse proceeds.
In the highest-resolution simulations, the Jeans length is resolved by at least 128 grid cells.
We call this number the ``Jeans parameter'', $N_{\rm J}$, as in \citet{HSC21, HSC22}. We note that previous work has shown that this is sufficient to resolve dynamo amplification during the collapse \citep{Sur10, Federrath11}.
When the peak densities reach $10^{-4} ~ \mathrm{g ~ cm^{-3}}$, the minimum spatial resolutions are $1.91\time 10^{-5}$~AU for $\gamma_{\rm eff} = 1.0$, $7.65\times 10^{-5}$~AU for $\gamma_{\rm eff}=1.1$, and $1.22\times 10^{-3}$~AU for $\gamma_{\rm eff} = 1.2$, respectively.
We also perform higher/lower-resolution runs, as a resolution study for the $\gamma_{\rm eff} = 1.1$ case, with $N_{\rm J}=$ 32, 64, and 256.

We terminate the simulations when the peak density, $\rho_{\rm peak}$, reaches $10^{-4} ~ \mathrm{g ~ cm^{-3}}$. 
We continue only the simulation with $\gamma_{\rm eff} = 1.1$ and $N_{\rm J}=128$ for another 2~orders of magnitude in higher density, because this is our main simulation model, which allows us to study the saturation phase of the magnetic field.
We use the \textsc{yt} toolkit \citep{yt}\footnote{\url{https://yt-project.org/}} to analyze the simulation data.

\section{Results}\label{sec:results}
When we analyze our simulation results, we use physical quantities only in a sphere with a radius of half the Jeans length, $L_{\rm J}$, as \cite{HSC21, HSC22} did.
Hereafter, we call this sphere ``Jeans volume''. 
Also, all averaged quantities in this paper are cell volume-weighted averages.

Here, we briefly describe the methodology to evaluate $L_{\rm J}$ in every data snapshot \citep[see also][]{HSC22}.
\begin{enumerate}
    \item Cut out an overdense region, with densities of $\rho \geq \rho_{\rm peak}/16$ centered on the position of the density maximum.
    
    \item Calculate the center of mass $\vb{r}_{\rm c}$ and `tentative' Jeans length $\left( = \sqrt{\pi c_{\rm s,t}^2/G\rho_{\rm mean,t}}\right)$ from the average density $\rho_{\rm mean, t}$, and sound speed $c_{\rm s,t}$ in the overdense region, and then assume half of this Jeans length to be the tentative core radius $r_0$.\label{it:temp_r0}
    
    \item Recalculate the tentative Jeans length and radius $r_1$ as done in step~\ref{it:temp_r0}, but now within the sphere of radius $\vb{r}_{\rm c}$ centered on $r_0$. \label{it:temp_r1}
    
    \item With the new cloud center and radius, go back to step~\ref{it:temp_r1}, and repeat it until $|r_1 - r_0|$ becomes smaller than the smallest cell width.
    
    \item Obtain $r_1$ as the core radius $r_1(=L_{\rm J}/2)$ after its value converges.
\end{enumerate}

\begin{figure}[htbp]
 \centering
  \plotone{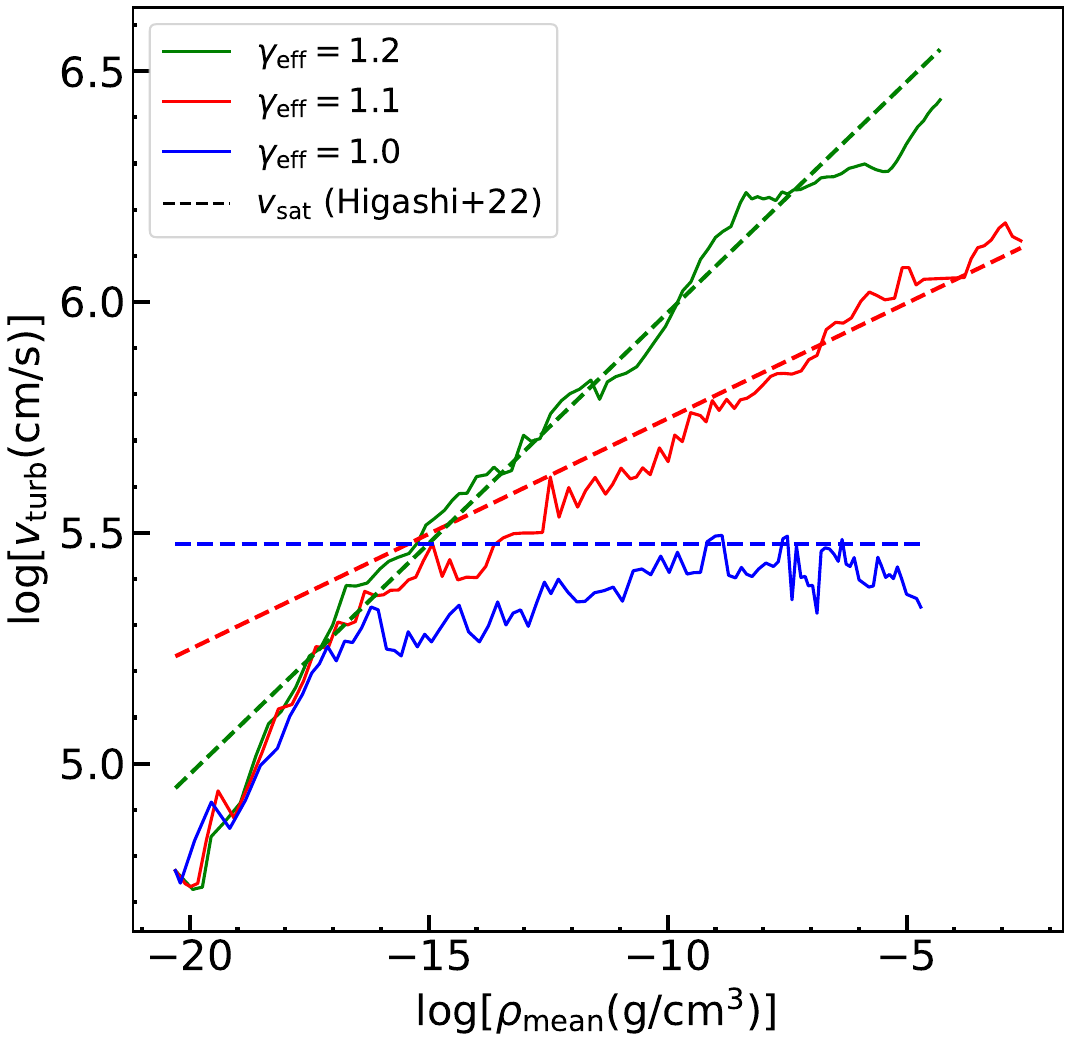}
  \caption{Evolution of the turbulent velocity $v_{\rm turb}$ as a function of mean density $\rho_{\rm mean}$ in the Jeans volume. The solid curves with different colors denote the simulation results for different $\gamma_{\rm eff}$. The dashed lines are the analytic estimates of the saturation velocities in \citet[eq.~17]{HSC22}.}
    \label{fig:velocity} 
\end{figure}

\subsection{Overall Evolution of the Fields}\label{ssec:overall}
Figure~\ref{fig:velocity} shows the evolution of the turbulent velocity as a function of mean density $\rho_{\rm mean}$ within the Jeans volume in each snapshot.
The solid curves denote the simulation results and the dashed lines indicate our analytic estimates presented by \citet[eq.~17]{HSC22}.

The turbulent velocity within the Jeans volume $V_{\rm J}$ is defined as \citep{HSC21, HSC22}
\begin{equation}
    v_{\rm turb}^2\equiv \sum_{\leq L_{\rm J}/2 } \frac{V_i}{V_{\rm J}}(\vb{v}_i-\vb{v}_{{\rm rad},i})^2, \label{eq:vturb}
\end{equation}
where $V_i, ~ \vb{v}_i, ~ \vb{v}_{\rm{rad},i}$ are the cell volume, the velocity, and smoothed radial velocity of the $i$th cell, respectively.
$\vb{v}_{{\rm rad}, i}$ is estimated as follows:
\begin{enumerate}
    \item Calculate the radial velocity profile within the Jeans volume with $N_{\rm rad}$ radial bins.
    To remove the fluctuations of the radial velocity around the scale of the cell size, $N_{\rm rad}$ should be smaller than the Jeans parameter so that we set $N_{\rm rad} = 16$.
    \item Interpolate the profile at the position of each cell to estimate the smoothed radial velocity.
\end{enumerate}
As we found in our previous (purely hydrodynamical) simulations \citep{HSC21,HSC22}, the turbulent velocities increase and saturate at the level of a few times the sound speed, also in the present MHD simulations.
Although the difference in growth rates among the three models is not 
clear because the dynamic range of the rapid increase regime is narrow, the growth rates are almost consistent with our analytic estimate in \citet[eq.~14]{HSC21}.
The reason for this is that the analytically estimated growth rates depend on $\gamma_{\rm eff}$, but their values are very similar.
In addition, the saturation levels of the turbulent velocities are also consistent with the analytic estimate in \citet[eq.~17]{HSC22}.

This means the effects of the back-reaction of the magnetic field onto the flow, the energy transportation between the magnetic field and the turbulent motion, have little impact on the growth and saturation of the turbulent velocity.

\begin{figure}[htbp]
 \centering
  \plotone{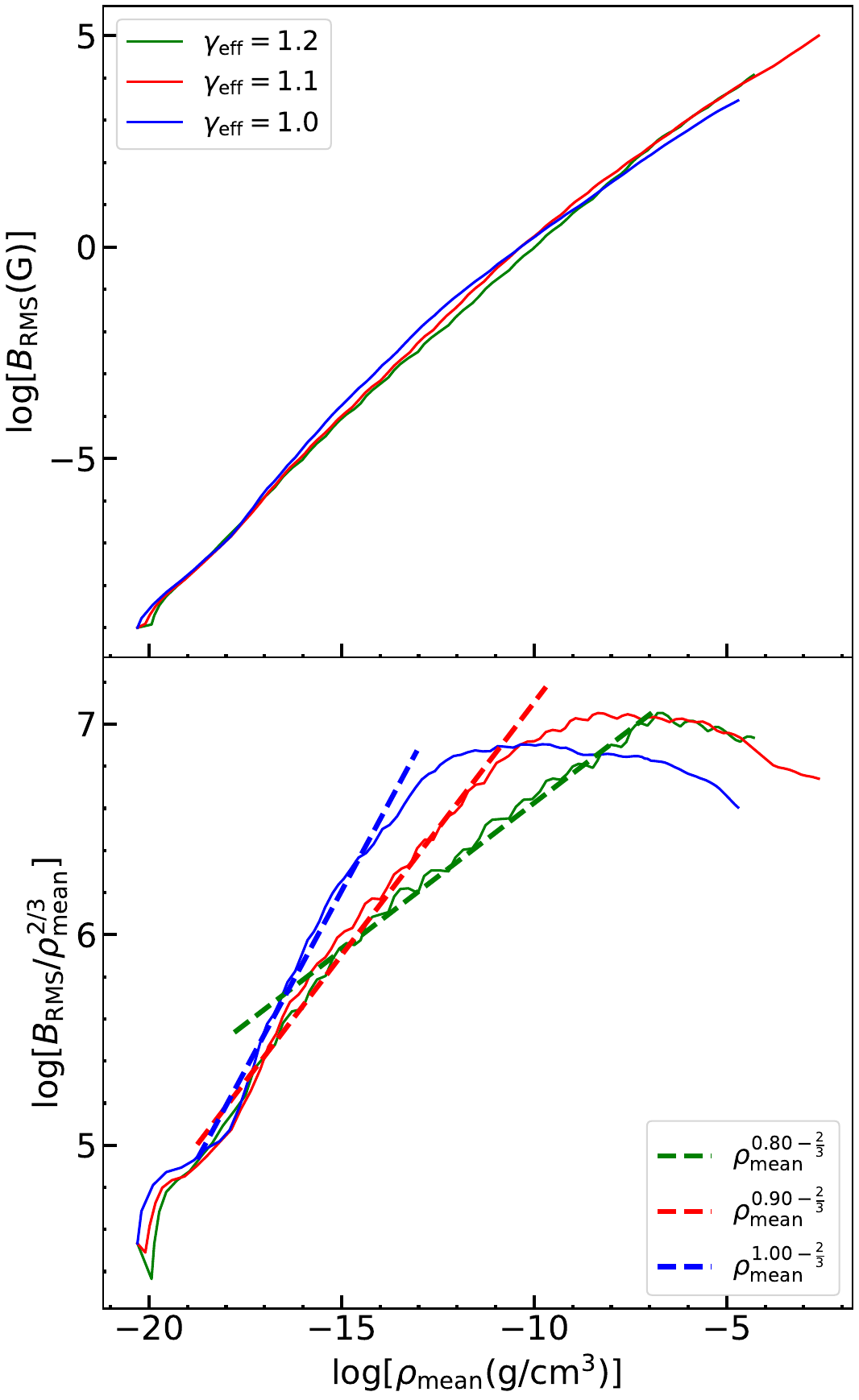}
  \caption{Upper panel: evolution of the RMS magnetic field strength $B_{\rm RMS}$ as a function of the mean density $\rho_{\rm mean}$ in the Jeans volume. Lower panel: magnetic field strength divided by $\rho_{\rm mean}^{2/3}$ to compensate the effect of the field growth due to flux freezing and contraction during the collapse. The solid curves with different colors denote the simulation results for different $\gamma_{\rm eff}$. The dashed lines show the approximated growth rate for each model.}
    \label{fig:b-field} 
\end{figure}

Figure~\ref{fig:b-field} shows the evolution of the root-mean-square (RMS) magnetic field strength $B_{\rm RMS}$ within the Jeans volume (upper panel) and its division by $\rho_{\rm mean}^{2/3}$ to compensate the effect of the amplification due to flux-freezing (bottom panel).
We see that the initially very weak ($=10^{-9}$~G) magnetic field increases during the collapse and is eventually amplified by a factor of $10^{12}-10^{15}$.
The field strength at the end of the collapse, especially for $\gamma_{\rm eff}=1.1$ (red curve) is consistent with the results of a previous study \citep{Sadanari23}.

In the lower panel, focusing on the primordial case, the growth rate is $B_{\rm RMS} \propto\rho^{0.9}$.
This is almost identical to the simulation result of cosmological simulations by \citealt{Turk2012} ($B\propto \rho^{0.89}$). 

On the other hand, comparing the 3~models, while the growth rates of the magnetic fields in the very early phase $(\rho_{\rm mean} \lesssim 10^{-17} ~ \mathrm{g~ cm^{-3}})$ are almost identical, they are different in the late stage, depending on $\gamma_{\rm eff}$. 
We will discuss the differences in the growth rates in this stage (kinematic stage) in detail, in later Sections~\ref{sec:estimate} and~\ref{sec:comparison}.

\subsection{Turbulence/Magnetic Field Spectra}\label{ssec:spectra}
To take a closer look at the detailed evolution of the velocity and magnetic fields,
we compute both fields in Fourier space (hereafter $k$-space) using a fast Fourier transform code. 
To make the analysis easier, we remap the simulation results onto a uniform grid with the same cell size as the highest refinement level.
We here note that, unlike our earlier work \cite{HSC21}, we cut out a cube with  $2 L_{\rm J}$ on each side to capture the emergence of the Kazantsev spectrum \citep[$\propto k^{3/2}$,][]{Kazantsev68,Kulsrud92} at low-$k$ end. This procedure is similar to the Fourier analysis method applied in \citet{Federrath11}.

We calculate the specific turbulent kinetic energy $\left(\varepsilon_{\rm turb} \right)$ and the specific magnetic energy $\left(\varepsilon_{\rm mag} \right)$ spectra as
\begin{eqnarray}
    \varepsilon_{\rm turb}&=&\frac{1}{2} \int \left|\hat{\vb{v}}_{\rm turb}(\vb{k})  \right|^2 4 \pi k^2 dk , \nonumber \\
    \varepsilon_{\rm mag}&=&  \frac{1}{2} \int \frac{\left|\hat{\vb{B}}(\vb{k}) \right|^2}{4\pi\rho_{\rm mean}} 4 \pi k^2dk,
\end{eqnarray}
where $\vb{\hat{v}_{\rm turb}(\vb{k}})$ and $\vb{\hat{B}(\vb{k}})$ are the turbulent velocity and the magnetic field in $k$-space, and $\vb{k}$ is the wave vector.

\begin{figure}[htbp]
 \centering
  \plotone{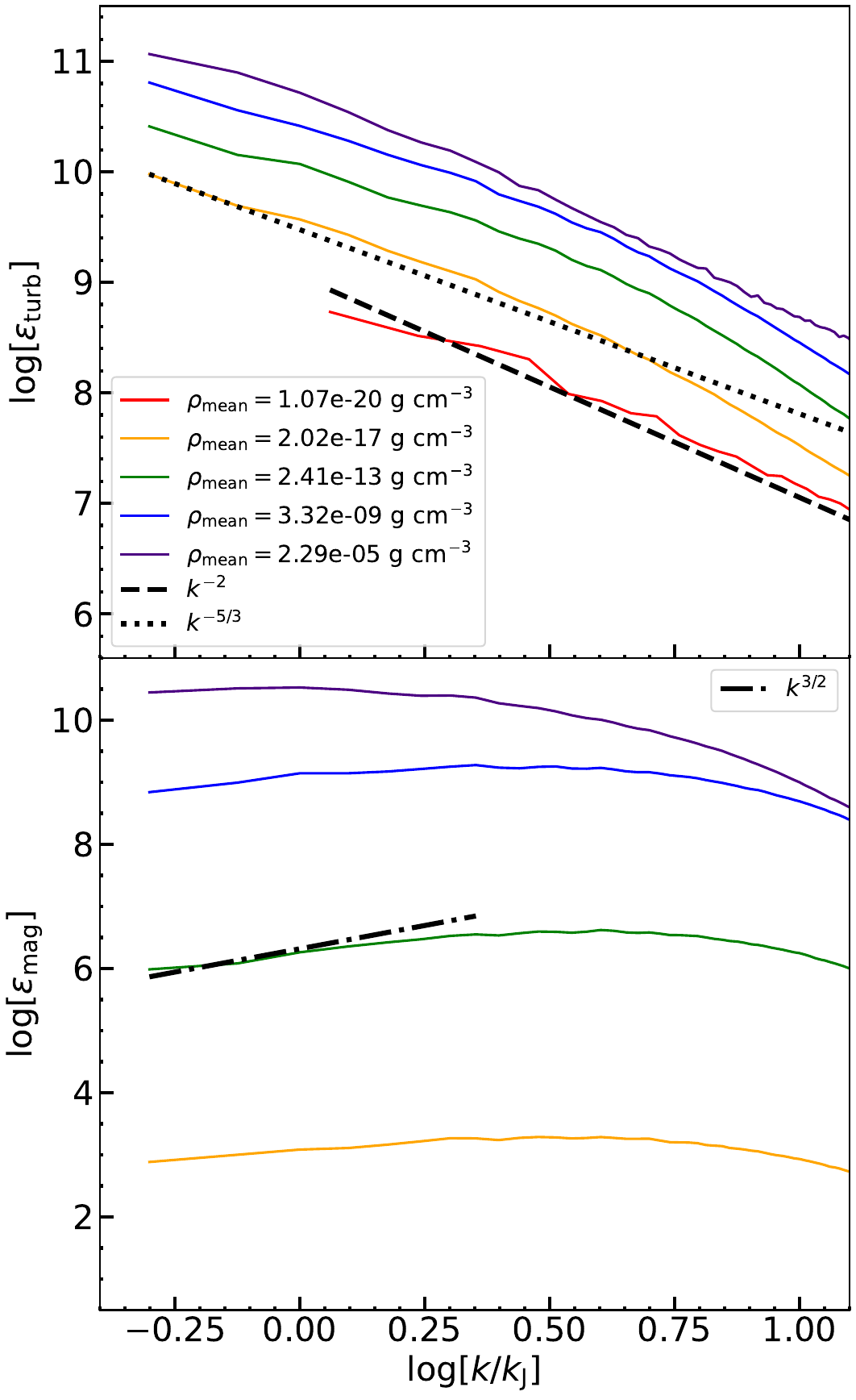}
  \caption{Evolution of the specific turbulent kinetic energy spectra (upper panel) and the specific magnetic energy spectra (lower panel), as a function of wavenumber $k$, scaled by the Jeans wavenubmer $k_{\rm J}$. 
    The different colors denote different $\rho_{\rm mean}$ to indicate the time evolution during the collapse.
    The black dashed, dotted, and dash-dotted lines denote $\propto k^{-5/3}$, $\propto k^{-2}$, $\propto k^{3/2}$, respectively.
}
    \label{fig:spectra} 
\end{figure}

We plot the obtained spectra for the $\gamma_{\rm eff}=1.1$ run in Figure~\ref{fig:spectra}.
The upper panel depicts the specific turbulent kinetic energy spectra, while the lower panel shows the specific magnetic energy spectra.
Each color denotes the snapshot corresponding to $\rho_{\rm mean}$.
We are interested in the scale around the Jeans length, so we normalize the wavenumber by $k_{\rm J}$, where $k_{\rm J}$ is the wavenumber at the Jeans length.

In the upper panel, we see that the specific turbulent kinetic energy spectra roughly keep the initial spectra $\propto k^{-2}$ (black dashed line) in the early phase $(\rho_{\rm mean} \lesssim 10^{-17} ~ \mathrm{g~cm^{-3}})$, as seen in \cite{HSC21}.
In the lower panel, the magnetic spectra have a mild peak at a smaller k (larger spatial scale) as collapse proceeds. 
This is good evidence that a small-scale dynamo is in action in the collapsing gas cloud.
In the kinematic regime, corresponding to the green and blue curves $(10^{-17} < \rho_{\rm mean} < 10^{-8} ~\mathrm{g~cm^{-3}})$, there is some evidence of a Kazantsev spectrum ($\propto k^{3/2}$; black dot-dashed line) emerging in the low-$k$ limit $(\log[k/k_{\rm J}] < 0.3)$.
At the same scale range, the turbulent kinetic energy is consistent with the Kolmogorov ($\propto k^{-5/3}$; black dotted line) spectrum.
These results are in agreement with \cite{Brandenburg05} and \cite{Federrath11}.
The simulation results also suggest that the spectrum is likely flatter than the Kazantsev one at the low-$k$ end, after saturation (purple curve) in this collapse simulation, as also shown in the uniform-grid simulation of \cite{Seta20}.
While one can suggest that the growth still continues at a lower $k$ than $k_{\rm J}$ after the saturation occurs at $k_{\rm J}$, we do not fully understand the reason for the flattening of the spectrum after the saturation.
We plan to investigate this further in a future study.

\subsection{Saturation Level of the Magnetic Field}\label{ssec:sat_level}
\begin{figure}[htbp]
 \centering
  \plotone{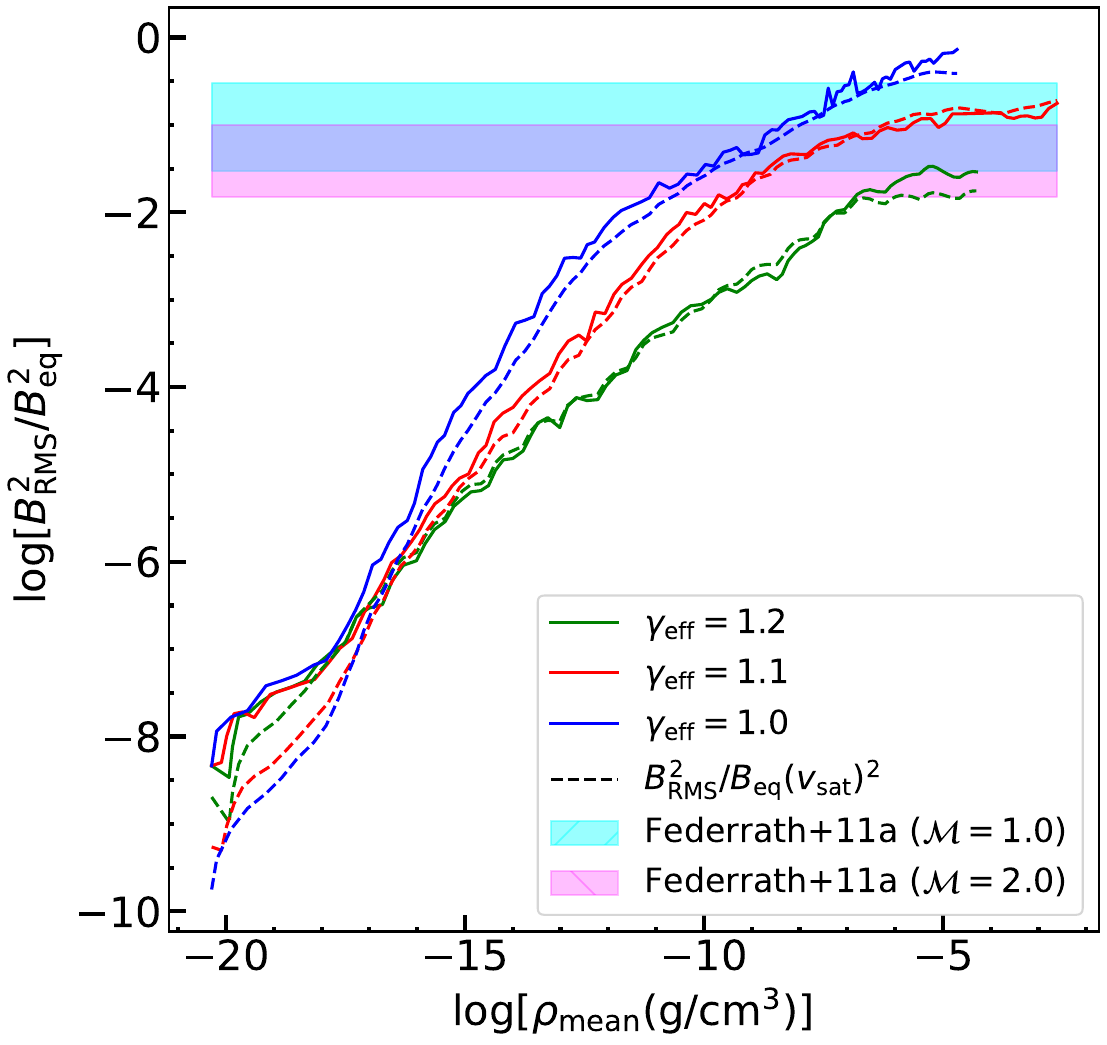}
  \caption{Evolution of the ratio between the specific magnetic energy and the specific turbulent kinetic energy. The solid curves are drawn by utilizing $v_{\rm turb}$ from simulation results shown in Figure \ref{fig:velocity}. The dashed curves are estimated using $v_{\rm sat}$ from \cite{HSC22}. The shaded areas denote the saturation levels obtained from the results of turbulence-in-a-box simulations with an isothermal EoS for driving Mach numbers $\mathcal{M} =1.0$ (cyan) and $2.0$ (magenta) in \cite{Federrath11_turb}.}
    \label{fig:bratio} 
\end{figure}

When the magnetic field strength amplified by the turbulence reaches a saturation level, the Lorentz force has become strong enough to counter the stretching, twisting, and folding actions induced by the turbulence during the exponential growth phase of the turbulent dynamo \citep{Federrath16}. The saturation level can maximally become as high as the equi-partition between the specific magnetic energy and the specific turbulent kinetic energy. However, the actual saturation level depends on the properties of the turbulence, in particular the sonic Mach number and the driving mode of the turbulence \citep{Federrath11_turb,AchikanathChirakkara21}.
To quantify the saturation level of the magnetic fields, we introduce the ratio $B_{\rm RMS}^2/B_{\rm eq}^2$ (see Figure~\ref{fig:bratio}).
$B_{\rm RMS}$ denotes the root mean square of the field strength in the Jeans volume.
$B_{\rm eq}$ is the equi-partition level of the magnetic field defined as
\begin{equation}
    B_{\rm eq} \equiv (4\pi \rho_{\rm mean})^{1/2} v_{\rm turb}.
\end{equation}

The solid curves are drawn by utilizing $v_{\rm turb}$ from the simulation results (solid curves in Figure \ref{fig:velocity}), while the dashed curves are estimated by using the analytic saturated turbulent velocities \citep{HSC22} (dashed lines in Figure~\ref{fig:velocity}).
As shown in Figure \ref{fig:velocity}, $v_{\rm turb}$ and $v_{\rm sat}$ are almost identical in the late phase, so that the two ratio are also almost identical.
The dashed curves are smoother than the solid ones by construction, which enables us to observe the clear saturation behavior.

For comparison, we plot the shaded areas, which denote the saturation levels obtained from turbulent-in-a-box simulations with an isothermal EoS for driving Mach numbers $\mathcal{M}=1.0$ (cyan) and 2.0 (magenta) in \cite{Federrath11_turb}.
The range of the shaded area is given by the difference in saturation level between turbulence driving modes, that is, compressive or solenoidal \citep{Federrath10}; the former gives an estimate for the lower limit of the expected saturation level, while the latter provides an estimate of the upper limit. The mixture of the two extreme driving modes corresponds to an intermediate value.

In Figure~\ref{fig:bratio}, we see that the saturation levels depend on $\gamma_{\rm eff}$.
For $\gamma_{\rm eff} = 1.1$ and $1.2$, it can be clearly seen that the specific energy ratio becomes constant at the end of both simulations, whereas it is unclear for $\gamma_{\rm eff}=1.0$. While a turnover to a likely saturation is underway in this model, the saturation process has not been completed.
These results indicate that the lower the $\gamma_{\rm eff}$, the higher the saturation level.
As shown in the upper panel of Figure~\ref{fig:b-field}, all of the magnetic field strengths are almost identical, whereas the turbulent velocities depend on $\gamma_{\rm eff}$.
This indicates that the differences in the saturation levels originate from the differences in the turbulent velocity among the three runs.
In summary, we find that the magnetic energies reach some fractions of the equi-partition level of about 3\% -- 100\% depending on $\gamma_{\rm eff}$, until the protostellar density $\left( \simeq 10^{-5} ~ \mathrm{g  ~ cm^{-3}} \right)$ is reached.
This shows that the magnetic field strength can amplify to a significant level in the collapsing phase, which will likely affect the dynamics of the core, and ultimately also the parental cloud.

\section{Analytic Estimate of the Growth rate}\label{sec:estimate}
In this section, we introduce analytic estimates for the growth rate of the magnetic field through gravitational contraction, based on the arguments of \cite{Xu20}, \cite{Mckee20}, and \cite{Stacy22}.

\subsection{Kinematic Dynamo Stage}\label{ssec:kinematic}
In the stage when the magnetic energy is much smaller than the kinetic energy, the magnetic field lines are stretched, twisted, and folded by turbulence on the viscous scale $\ell_\nu$ (i.e., smallest eddy scale), where the subscript $\nu$ labels the quantities at the viscous scale.
This stage is called the kinematic dynamo stage, and then the magnetic field is expected to be amplified with the exponential growth rate, $\Gamma$, as
\begin{equation}
    B^2 \propto \exp\left(\int \Gamma dt\right).\label{eq:Bsq_th}
\end{equation}

We assume that the turbulent velocity on the viscous scale $\ell_\nu$ is given by the extrapolation from a certain scale $L$ with a characteristic scaling law, that is, 
\begin{equation}
    v_\nu = v_{\rm L}\left(\frac{\ell_\nu}{L}\right)^\vartheta, \label{eq:v_nu}
\end{equation}
where $L$ is the characteristic scale of the turbulence, which is typically of the order of the Jeans length $L_{\rm J}$ in collapsing gas clouds, and $\vartheta$ is in the range from 1/3 (Kolmogorov turbulence, \citealt{Kolmogorov41}) to 1/2 (compressible turbulence, \citealt{BURGERS1948}); see also the discussion and simulations in \citet{Federrath21}.
The growth rate of the magnetic fields, $\Gamma_\nu$, at the viscous scale \citep{Schober12, Bovino13} is given as
\begin{equation}
    \Gamma_\nu = \frac{v_\nu}{\ell_\nu} = \left(\frac{v_{\rm 
    turb}}{L_{\rm J}}\right)\mathrm{Re}^{(1-\vartheta)/(1+\vartheta)},\label{eq:Gamma_nu}
\end{equation}
where 
$\mathrm{Re}$ is the Reynolds number \citep{Haugen04, Federrath11} defined as
\begin{equation}
    \mathrm{Re} = L_{\rm J} v_{\rm turb}/\nu \simeq N_{\rm J}^{4/3}.
\end{equation}

Additionally, in collapsing primordial gas clouds, a high ionization degree enables the condition in which the gas and magnetic flux are fully coupled at the Jeans scale \citep{Maki&Susa2004,Maki07,Susa2015, nakauchi19}.
For isotropic, uniform collapse with this condition, i.e., the magnetic field is randomly oriented in the core, the magnetic field will increase as $B \propto \rho^{2/3}$ even without the dynamo growth, i.e., purely due to flux conservation during collapse.
Thus, the sum of the growth effects due to flux-freezing and kinematic dynamo gives
\begin{equation}
    B_{\rm kin}^2 = B_0^2 \xi^{4/3} \exp\left(C_{\Gamma}\int\Gamma_{\nu}dt\right),\label{eq:B_kinematic1}
\end{equation}
where $B_0$ denotes the initial field strength, and $\xi$ is the ratio between the density $\rho$ and the initial density $\rho_0$.
In the above equation, $C_{\Gamma}$ is a constant, introduced by \cite{Stacy22}, possibly determined by the properties of turbulence, such as the driving mode mixture and the magnetic Plandtl number \citep{Federrath11_turb, Federrath14}.
In \cite{Stacy22}, they adopt $C_{\Gamma}=3/4$, whereas we assume $C_{\Gamma}=0.075$.
The reason of this choice is discussed in Section \ref{sec:comparison}.
From Equation~(\ref{eq:Gamma_nu}), assuming the outer scale of turbulence as $L_{\rm J}$, we have  
\begin{equation}
    \int\Gamma_{\nu}dt = \left(\frac{3}{32}\right)^{1/2}\mathrm{Re}^{(1-\vartheta)/(1+\vartheta)}\langle\mathcal{M}_{\rm turb}\rangle \tau, \label{eq:B_kinematic2}
\end{equation}
where $\langle\mathcal{M}_{\rm turb}\rangle$ is the time-averaged turbulent Mach number in the region where $\rho_{\rm mean} \leq 10^{-12} ~ \mathrm{g ~ cm^{-3}}$.
In the above equation, we use the relation
\begin{equation}
    L_{\rm J} = \left(\frac{\pi c_{\rm s}^2}{G\rho}\right)^{1/2}=\left(\frac{32}{3}\right)^{1/2}c_{\rm s}t_{\rm ff} \label{eq:LJ_tff}    
\end{equation}
We also define 
\begin{equation}
    \tau\equiv \int (1/t_{\rm ff}) dt,
\end{equation}as in \cite{Sur10} and \citet{Federrath11}.

The magnetic field keeps growing exponentially due to the kinematic dynamo until the magnetic energy becomes comparable to the turbulent kinetic energy at the viscous scale.

\subsection{Non-linear Dynamo Stage}\label{ssec:nonlinear}
When the magnetic energy becomes comparable to the turbulent kinetic energy at the viscous scale, the suppression of the turbulent dynamo action kicks in.
After the suppression of the growth at the smaller scale, the larger scale eddies take over the dynamo growth, until the two energies are comparable at that scale.
It is theoretically expected that this cycle continues until the equipartition is achieved up to the driving scale of the turbulence.\footnote{However, the equipartition level at the driving scale is usually lower than unity, depending on the magneto-hydrodynamic properties. For example, \cite{Federrath11_turb, Federrath14} indicated that the equipartition level depends on the driving mode, the Mach number, and the Plandtl number of the turbulence.}

This stage is called the non-linear dynamo stage.
In this stage, the treatment of the kinematic stage, where only the microscopic diffusion exists and the magnetic flux is frozen-in on larger scales than the dissipation scale of the magnetic energy, is no longer valid.
The developed MHD turbulence operates the fast turbulent reconnection and magnetic flux diffusion becomes a non-negligible effect \citep[see][for a review in detail]{Lazarian15}.
As a result, the flux-freezing condition is violated, and the growth rate becomes smaller \citep{Xu16, Xu20}.

\cite{Xu20} derived the equation for a non-linear dynamo in a time-dependent medium with a self-consistent treatment, and \cite{Stacy22} extended their treatment to describe the evolution of the dynamo up to the late stage of the collapse phase.
However, both studies assumed an isothermal EoS,
even though the gas is not necessarily isothermal in the primordial environment.
For example, in \cite{Stacy22}, the thermal evolution for  a number density $\gtrsim 10^{10}{\rm cm^{-3}}$ is almost isothermal (see fig.~11 in their paper), but it is not in \cite{Yoshida08}, at the same density range.
In fact, the detailed thermal evolution shown in different works in the literature slightly differs from each other, depending on the implemented chemistry network and radiation transfer prescription.
Thus, we further generalize their arguments by considering the dependence on $\gamma_{\rm eff}$.

In the non-linear stage, the dynamo growth can occur at $k < k_{\rm p}$, where $k_{\rm p}$ denotes the wavenumber at the kinetic-magnetic energy equipartition is achieved.
We obtain the time-dependent specific magnetic energy $\varepsilon_{\rm B}$ by the integral form of the Kazantsev spectrum \citep{Kazantsev68, Kulsrud92}
as,
\begin{equation}
    \varepsilon_{\rm B} = \varepsilon_{\rm B, ref} \left(\frac{k_{\rm p}}{k_{\rm ref}}\right)^{5/2}
    \exp\left(\frac{3}{4}\int_{t_1}^{t}\Gamma dt\right), \label{eq:epsilon_B1_1}
\end{equation}
where the quantities with the subscript `ref' denote the reference values in the contracting system, for example, the time-varying core radius of the collapsing cloud.
The quantities with the subscript `1' denote the values at the onset of the non-linear stage.
The flux-freezing condition gives the growth of the reference specific magnetic energy as 
$\varepsilon_{\rm B, ref}=\varepsilon_{\rm B1}\left(\xi/\xi_1\right)^{1/3}$.
The reference wavenumber $k_{\rm ref}$ scales inversely with the length as the contraction proceeds.
Considering that the reference scale is the Jeans length, it is given as $k_{\rm ref} \propto L_{\rm J}^{-1} \propto \xi^{1-\gamma_{\rm eff}/2}$.
So, we obtain
\begin{equation}
      k_{\rm ref} = \left(\xi/\xi_1\right)^{1-\gamma_{\rm eff}/2}k_{\rm ref,1}\label{eq:kref}.
\end{equation}
On the other hand, using the scaling law in Equation (\ref{eq:v_nu}) for the turbulent velocity, we obtain 
$v_{\rm p} = v_{\rm turb}\left(k_{\rm ref}/k_{\rm p}\right)^{\vartheta}$.
The magnetic energy at $k_{\rm p}$ is given by the turbulent kinetic energy at the same scale because the energy equipartition is established for $k > k_{\rm p}$:
\begin{equation}
    \varepsilon_{\rm B} = \frac{1}{2}v_{\rm p}^2 = \frac{1}{2}v_{\rm turb}^2 \left(\frac{k_{\rm ref}}{k_{\rm p}}\right)^{2\vartheta}.\label{eq:epsilon_B1_2}
\end{equation}
The eddy turnover rate at $k_{\rm p}$ is
\begin{equation}
    \Gamma = \Gamma_{\rm p} = v_{\rm p}k_{\rm p}. 
\end{equation}
We note that we consider only $\vartheta=1/3$ in the non-linear stage, unlike the kinematic stage.
The reason of this is discussed later (Section~\ref{sec:comparison}).

By combining the above two equations and assuming the turbulent velocity is saturated (i.e., $v_{\rm turb}=v_{\rm sat}$) at the onset of the non-linear stage, we obtain 
\begin{equation}
    \Gamma \varepsilon_{\rm B} = \frac{1}{2}v_{\rm sat}^3 k_{\rm ref}\label{eq:gamma_epsilon1}
\end{equation}
Considering the density dependence of $k_{\rm ref}$ and $v_{\rm sat}\propto c_{\rm s}$, the above equation can be rewritten as
\begin{equation}
    \Gamma \varepsilon_{\rm B} =\frac{1}{2}\varepsilon_1 \left(\frac{\xi}{\xi_1}\right)^{\gamma_{\rm eff} -1/2},\label{eq:gamma_epsilon2}
\end{equation}
where $\varepsilon_1 = v_{\rm sat, 1}^3k_{\rm ref,1}$. 

Applying $d\ln/ dt$ to both sides of Equations (\ref{eq:epsilon_B1_1}) and (\ref{eq:epsilon_B1_2}), vanishing $k_{\rm p}$ by combining them, and using Equation (\ref{eq:gamma_epsilon2}), we find
\begin{equation}
    \frac{d\varepsilon_{\rm B}}{dt} = \varepsilon_{\rm B}
    a_\gamma
    \frac{d}{dt}\left[\ln\left(\frac{\xi}{\xi_1}\right)\right]
    +\chi \varepsilon_1 \left(\frac{\xi}{\xi_1}\right)^{\beta_{\gamma}}\label{eq:depsilon_dt},
\end{equation}
where
\begin{eqnarray}
    a_{\gamma} &=& \frac{15}{19}\gamma_{\rm eff}- \frac{41}{57},\nonumber \\
    \beta_{\gamma}&=&\gamma_{\rm eff} -1/2,\nonumber \\
    \chi &=& \frac{3}{38}.\label{eq:a_beta_chi}
\end{eqnarray}
We finally obtain the following equation by integrating Equation~(\ref{eq:depsilon_dt}) in the same manner as in \cite{Stacy22}:
\begin{equation}
    \varepsilon_{\rm B} = \left(\frac{\xi}{\xi_1}\right)^{a_{\gamma}}\varepsilon_{\rm B,1}+\frac{\chi\varepsilon_1 \xi^{a_{\gamma}}}{\xi_1^{\beta_{\gamma}}}\int_{t_1}^{t}\xi(t')^{\beta_{\gamma}-a_{\gamma}}dt'. \label{eq:obtained_epsilon}
\end{equation}
The first term in Equation (\ref{eq:obtained_epsilon}) expresses the growth due to cloud collapse including the effects of turbulent reconnection diffusion, and the second is for the dynamo growth.
Assuming $\gamma_{\rm eff} =1$, Equation (\ref{eq:obtained_epsilon}) recovers the identical expression in \cite{Xu20} and \cite{Stacy22}.

In order to estimate the field strength, we use the same approximation as in \cite{Stacy22}.
Using eq.~(B17) in \cite{Mckee20}, we have
\begin{eqnarray}
    \int_{t_1}^{t}\xi(t')^{\beta_\gamma-a_{\gamma}}&dt'& \simeq \frac{2\phi_{\rm ff}t_{\rm ff,0}}{3\pi (\frac{1}{2}-\beta_\gamma+ a_\gamma)}\nonumber \\
    &\times& \left(\frac{1}{\xi_1^{\frac{1}{2}-\beta_\gamma + a_{\gamma}}}-\frac{1}{\xi^{\frac{1}{2}-\beta_\gamma + a_{\gamma}}}\right),
\end{eqnarray}
where $t_{\rm ff,0}=\xi^{1/2}t_{\rm ff}$ is the free-fall time at the initial density and $\phi_{\rm ff}$ is the ratio between the times that the gas collapses to a star $t_{\rm coll}(\rho)$ and the free-fall time $t_{\rm ff}(\rho)$.
This integration is correct if $\beta_\gamma - a_\gamma < 1/2$ is satisfied, which is true in the present case (c.f., Equations~\ref{eq:a_beta_chi}).
Substituting the above expression into Equation~(\ref{eq:obtained_epsilon}) and arranging the second term with Equation~(\ref{eq:LJ_tff}), we obtain
\begin{eqnarray}
    \varepsilon_{\rm B} &=& \left(\frac{\xi}{\xi_1}\right)^{a_{\gamma}}\varepsilon_{\rm B,1}
    +\frac{\chi\phi_{\rm ff}\mathcal{M}_{\rm sat}v_{\rm sat,1}^2}{2\sqrt{6}\pi \left(\frac{1}{2}-\beta_\gamma+a_{\gamma}\right)}\nonumber\\
    &\times &\left(k_{\rm ref,1}L_{\rm J,1}\right)
    \left[\left(\frac{\xi}{\xi_1}\right)^{a_{\gamma}}-\left(\frac{\xi}{\xi_1}\right)^{\beta_\gamma - 1/2} \right], \label{eq:obtained_epsilon_mod}
\end{eqnarray}
where $\mathcal{M}_{\rm sat}(\equiv v_{\rm sat,1 }/c_{\rm s,1})$ is the saturation Mach number in \citet[eq.~18]{HSC22}.
In the above equation, all density-dependent quantities are converted to the form of $Y(\xi)=Y_1(\xi/\xi_1)^{Z}$.
We also arrange the equation as
\begin{equation}
    \varepsilon_{\rm B} = \left(\frac{\xi}{\xi_1}\right)^{a_{\gamma}}\varepsilon_{\rm B,1}\mathcal{A}_{ \gamma, 1}^2,
\end{equation}
where
\begin{eqnarray}
    \mathcal{A}_{\gamma, 1}^2 &=& 1 + \frac{\chi\phi_{\rm ff}\mathcal{M}_{\rm sat}}{\sqrt{6}\pi \left(\frac{1}{2}-\beta_\gamma+a_{\gamma}\right)}\left(\frac{\frac{1}{2} v_{\rm sat,1}^2}{\varepsilon_{\rm B,1}}\right)\nonumber\\
    &\times& \left[1-\left(\frac{\xi}{\xi_1}\right)^{\beta_\gamma - 1/2-a_\gamma} \right].\label{eq:A_gamma_sq}
\end{eqnarray}
Here, we utilize the relation $k_{\rm ref,1}L_{\rm J,1}=1$, obtained by Equation~(\ref{eq:kref}). 
Unless the specific magnetic energy ($\varepsilon_{\rm B,1}$) is much smaller than the specific turbulent kinetic energy ($v_{{\rm sat},1}^2/2$), the contribution due to dynamo amplification is not significant (i.e., $\mathcal{A}_{\gamma, 1} \sim 1$).

Rewriting Equation (\ref{eq:obtained_epsilon_mod}), we have the analytic estimate of the magnetic field strength in the non-linear stage,
\begin{equation}
    B_{\rm nl} = B_1 \mathcal{A}_{\gamma,1}\left(\frac{\xi}{\xi_1}\right)^{(1+a_{\gamma})/2}. \label{eq:B_nonlinear}
\end{equation}
Finally, we need to assess $\xi_1$ to distinguish the two stages.
We can complete this task with the argument of the energy ratio.
With the definition for the beginning of the non-linear stage, the specific energies $\varepsilon_\nu$ of the magnetic field and the turbulence are identical at the viscous scale.
Assuming the Kazantsev spectrum ($\propto k^{3/2}$) for the specific magnetic energy and the Kolmogorov spectrum ($\propto k^{-5/3}$) for the specific turbulent kinetic energy, we can define the energy ratio at the onset of the non-linear stage $R_{\rm nl,1}$ as
\begin{equation}
    R_{\rm nl,1} = \frac{\int_{k_{\rm J}}^{k_\nu} \varepsilon_\nu (k/k_\nu)^{3/2} dk }{\int_{k_{\rm J}}^{k_\nu} \varepsilon_\nu (k/k_\nu)^{-5/3} dk }.
\end{equation}
Normalizing the wavenumbers by $k_{\rm J}$, the above equation can be written as
\begin{eqnarray}
    R_{\rm nl,1} &=& N_{\rm J}^{-19/6}\frac{\int_{1}^{N_{\rm J}} N^{3/2} dN }{\int_{1}^{N_{\rm J}} N^{-5/3} dN } \nonumber\\
    &=& \frac{4}{15}N_{\rm J}^{-19/6} \frac{N_{\rm J}^{5/2}-1}{1 - N_{\rm J}^{-2/3}},
    \label{eq:R_nl0}
\end{eqnarray}
where $k/k_{\rm J}=N$ and $k_{\nu} / k_{\rm J}= N_{\rm J}$.
For $N_{\rm J}=128$, Equation~(\ref{eq:R_nl0}) gives us $R_{\rm nl,1} \simeq 0.011$.
Hence we can evaluate $\xi_1$ as $\xi$ such that $\varepsilon_{\rm mag}/\varepsilon_{\rm turb}=R_{\rm nl,1}$ is satisfied, i.e., $\xi_1 = \xi\left(\varepsilon_{\rm mag}/\varepsilon_{\rm turb}=R_{\rm nl,1}\right)$.

\subsection{Equipartition} \label{ssec:equipartition}
When the specific magnetic energy reaches a certain fraction of the specific turbulent kinetic energy at the driving scale of the turbulence, the field growth due to the non-linear dynamo ends.
Afterward, the magnetic field strength can be estimated solely from the specific turbulent kinetic energy.
In addition, the turbulent velocity at the late phase of collapse is generally saturated and given by a few times the sound speed \citep{HSC22}.
Thus, the dependence of the equipartition field strength is 
\begin{equation}
    B_{\rm eq} \propto \left(4\pi\rho\right)^{1/2}v_{\rm sat}\propto \xi^{1/2}c_{\rm s}\propto \xi^{\gamma_{\rm eff}/2}. \label{eq:equi_xi}
\end{equation}

For comparison, we summarize the growth rates of the magnetic field in the form of $B\propto \xi^X$ in the non-linear stage without dynamo growth and the equipartition stage in Table~\ref{tab:growth_rate}.
\begin{table}[htb]
 \caption{Growth rates of the magnetic field $\left(B\propto \xi^X\right)$ in the non-linear stage without dynamo growth and the equipartition stage.}
  \centering
   \begin{tabular}{l|cc}
     \hline \hline
     $\gamma_{\rm eff}$ & Non-linear  & Equipartition  \\
     \hline
     1.0 & $ \xi^{0.535}$  & $ \xi^{0.500}$ \\ 
     1.1 &  $  \xi^{0.575}$  &$ \xi^{0.550}$\\
     1.2 & $\xi^{0.614}$  & $ \xi^{0.600}$\\
     \hline    
   \end{tabular}\\
  \label{tab:growth_rate}
\end{table}

\section{Comparison between Simulation Results and Analytic Estimates}\label{sec:comparison}

\begin{figure}[htbp]
 \centering
  \plotone{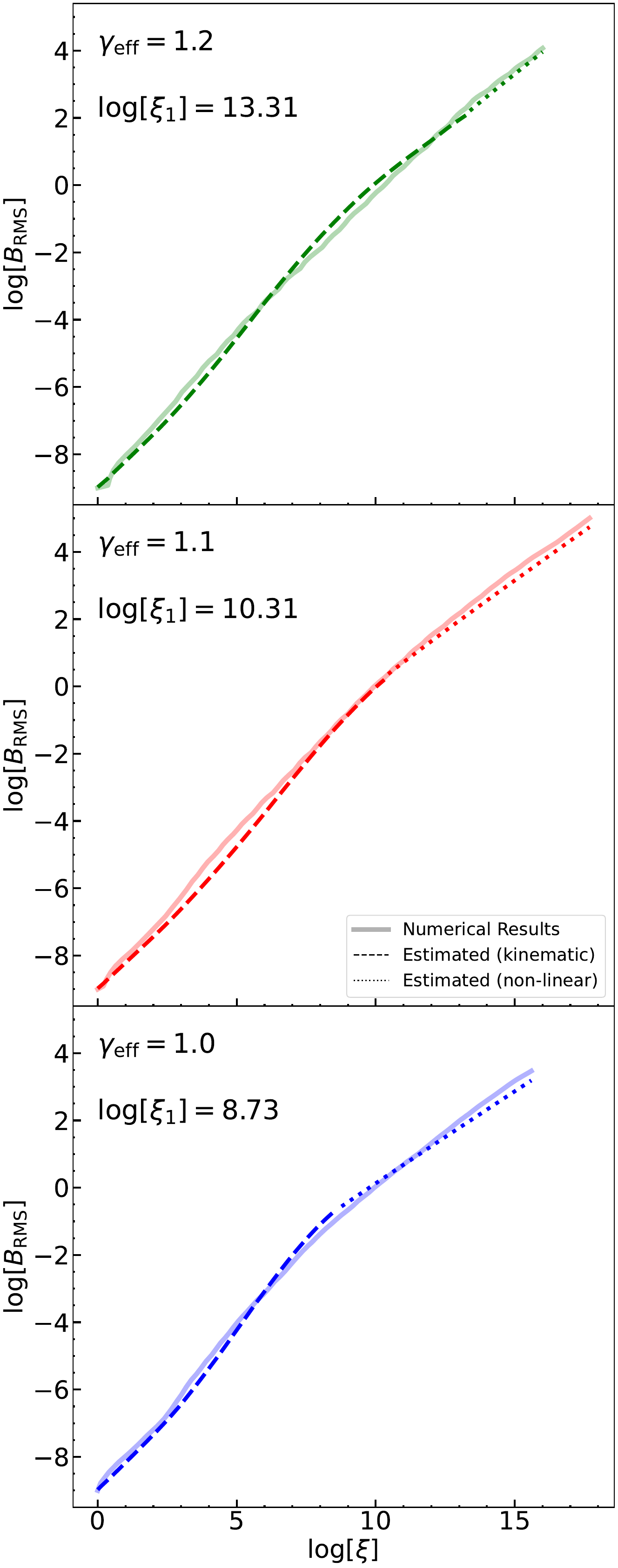}
  \caption{Comparison between the simulation results and the analytic estimates.
           From top to bottom, the panels depict the comparison for $\gamma_{\rm eff}=1.2$, 1.1, and 1.0, respectively.
           The $\xi_1$ values (separating the kinematic from the non-linear stage) are shown in the upper left in each panel.
           The solid curves are the simulation results.
           The dashed curves are obtained by Equation~(\ref{eq:B_kinematic1}).
           The dotted curves are obtained by Equation~(\ref{eq:B_nonlinear}).
           }
    \label{fig:b_comparison} 
\end{figure}

First, we compare our simulation results with the analytic values obtained by Equation~(\ref{eq:B_kinematic1}) and Equation~(\ref{eq:B_nonlinear}) in Figure~\ref{fig:b_comparison}.
From top to bottom, the panels depict the comparison for $\gamma_{\rm eff}=1.2$, 1.1, and 1.0, respectively.
The threshold $\xi_1$ shown at the upper left in each panel is calculated using Equation~(\ref{eq:R_nl0}).

\begin{figure}[htbp]
 \centering
  \plotone{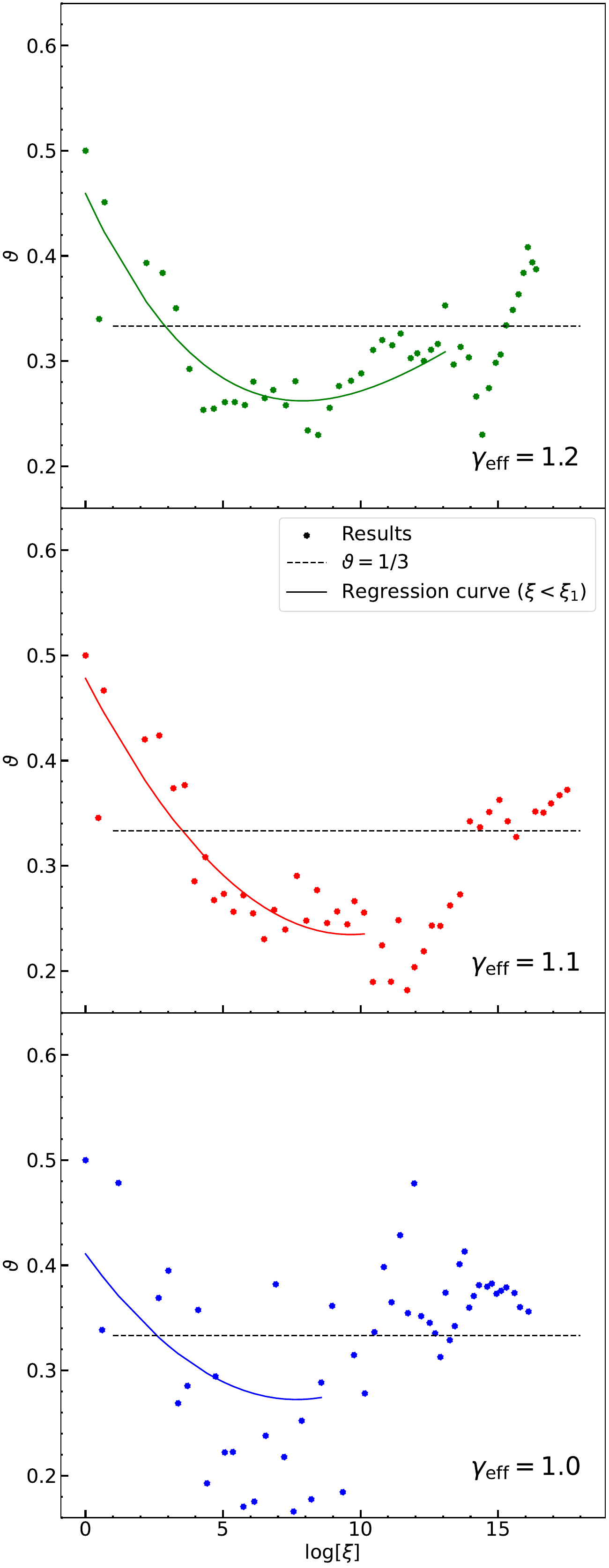}
  \caption{Scatter plot of the calculated turbulent velocity spectral indices for $\gamma_{\rm eff}=$1.2 (top), 1.1 (middle), and 1.0  (bottom).
           The solid curves are a quadratic fit in the range $\xi < \xi_1$ by using the least-square method.
           The dashed lines denote $\vartheta=1/3$.
           }
    \label{fig:index} 
\end{figure}

In order to measure the spectral indices $\vartheta$ in Equation~(\ref{eq:B_kinematic2}), we fit the simulation results. 
In the kinematic stage, the dynamo growth occurs at $k_\nu$, while it occurs at $k_{\rm p}$ in the non-linear stage.
In order to assess them from simulation results, we adopt the relation $k_\nu = 0.025 k_{\rm turb}\mathrm{Re}^{3/4}$ from \cite{Kriel22}, assuming $k_{\rm turb}/k_{\rm J}=1$.
$k_{\rm p}$ is the wavenumber where $\varepsilon_{\rm mag}$ is the maximum.
As a result, the estimated spectral indices $\vartheta$ are shown in Figure~\ref{fig:index}.
We also fit these points in the range $\xi < \xi_1$ by a quadratic function using the least-square method and plot them as solid curves.
\footnote{Note that the fitting by a quadratic function can be valid only in the kinematic stage for $\xi_1 \ll \infty$. For $\xi_1 \rightarrow \infty$, we might need another fitting function.}
Overall, the spectral indices (initially $\vartheta=1/2$) gradually decrease as the collapse proceeds and roughly decline to $\vartheta\sim1/3$ (dotted lines) in all models.
This is the reason why we assume $\vartheta=1/3$ in Section \ref{ssec:nonlinear}.

Going back to Figure~\ref{fig:b_comparison} and focusing on the dashed curves, these curves are obtained by using the fitting function in Figure~\ref{fig:index}.
We can see that the magnetic fields in the kinematic stage for all models are well reproduced by Equation~(\ref{eq:B_kinematic1}).
This result also means that the growth of the magnetic fields sensitively depends on the spectral index of the turbulent kinetic energy (see also \citet{Schober12}).

As shown in Section~\ref{ssec:kinematic}, we set $C_\Gamma=0.075$.
This is the best-fit parameter chosen for reproducing the simulation results, but possibly determined by the property of turbulence.
To clarify the physical dependence of $C_\Gamma$ is interesting, but is beyond the scope of this work.
We will address this issue in a future study.

In the non-linear stage, the remaining unknown is the value of $\phi_{\rm ff}$ in Equation~(\ref{eq:obtained_epsilon_mod}).
In \citet{Stacy22}, they assumed $\phi_{\rm ff}$ as a constant and then calculated its value by integrating the infall velocity of the gas.
In this work, for simplicity, we assume the cloud collapses in a self-similar fashion, which can be a good approximation of the ideal collapsing cloud as confirmed by \citet{HSC22}.
While they studied the non-MHD case, the initial magnetic field strength is very weak in the present work, and hence the same approximation holds here, i.e., the collapse dynamics are not much different from the non-magnetized collapse.
Since the gas core density is given by the collapse time $t_{\rm ss}$ as $\rho_{\rm mean} = \alpha(0)/4\pi G t_{\rm ss}^2$,  where $\alpha(0)$ is a constant depending on $\gamma_{\rm eff}$, $\phi_{\rm ff}$ can be analytically estimated as
\begin{equation}
    \phi_{\rm ff} = t_{\rm ss}/t_{\rm ff} = \left(\frac{32\alpha(0)}{12\pi^2}\right)^{1/2}.
\end{equation}
From \citet{HSC22}, $\alpha(0)=\left[1.67, 3.46, 7.19\right]$ for $\gamma_{\rm eff}=\left[1.0, 1.1, 1.2 \right]$, so we have $\phi_{\rm ff}= \left[0.67, 0.94, 1.39\right]$.

Focusing on the dotted curves in Figure~\ref{fig:b_comparison}, we see that the magnetic fields in the non-linear stage for all models can be reproduced to within a factor of 2.05 by our extended estimate based on Equation~(\ref{eq:B_nonlinear}).

To summarize, the overall magnetic field evolution can be reproduced by our extended analytic estimates (Equations~(\ref{eq:B_kinematic1}) and~(\ref{eq:B_nonlinear})) for various different values of $\gamma_{\rm eff}$.
The transition between the kinematic stage and the non-linear stage can also be estimated analytically from Equation~(\ref{eq:R_nl0}) using the numerical resolution.

\section{Resolution Study}\label{sec:resolution}
\begin{figure}[htbp]
 \centering
  \plotone{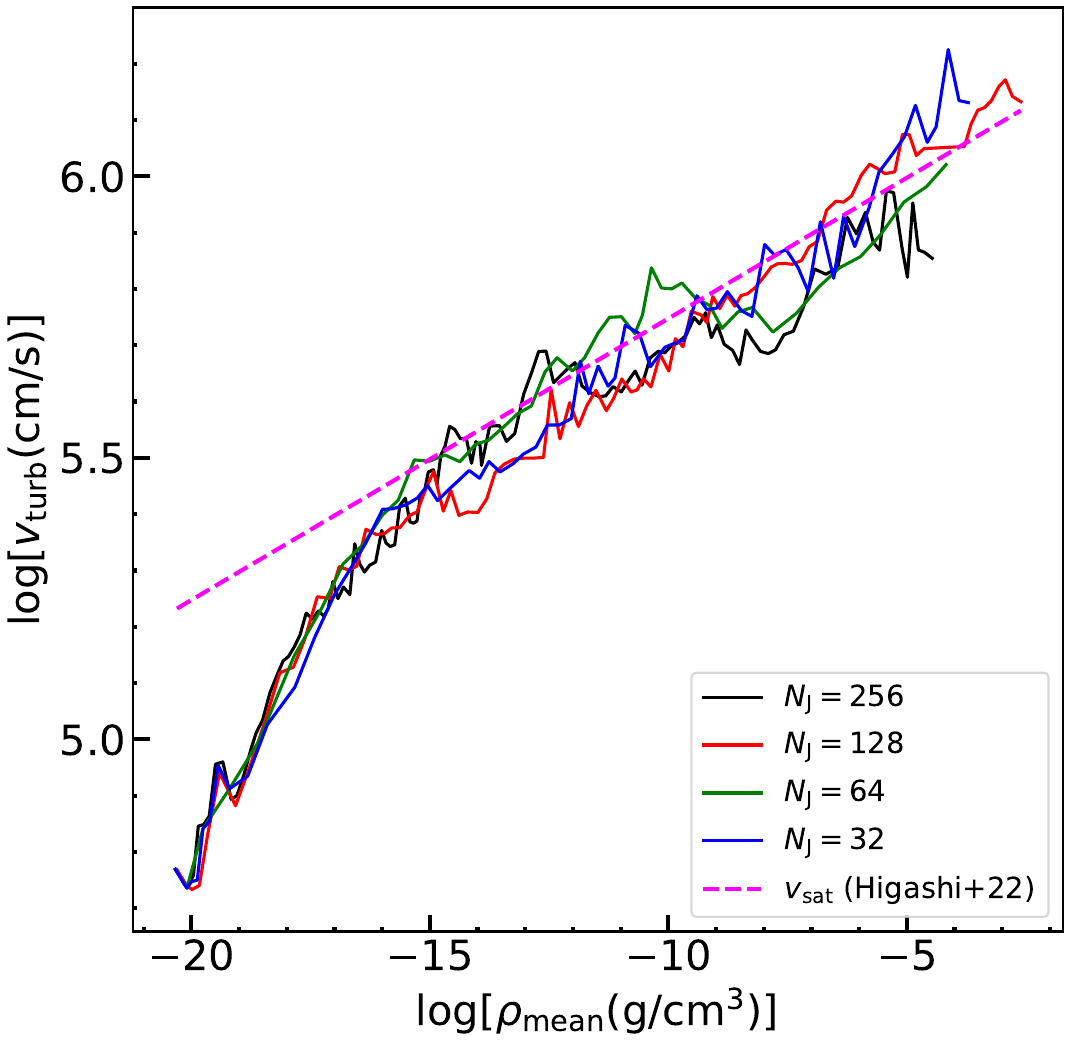}
  \caption{The same as Figure~\ref{fig:velocity}, but for different numerical Jeans resolutions.
  The different colors denote different numbers of grid cells to resolve the Jeans length.
  }
    \label{fig:velocity_reso} 
\end{figure}

In this section, we present the results of the resolution study for $\gamma_{\rm eff}=1.1$ to reinforce the arguments in the earlier sections.
First of all, we compare the turbulent velocities for different numerical resolutions (Jeans resolution $N_{\rm J}$) in Figure~\ref{fig:velocity_reso}.
As shown in Section~\ref{sec:method}, $N_{\rm J}$ is the number of grid cells per Jeans length , i.e., this number doubles when the minimum cell width is halved.

We find that the turbulent velocity converges to the saturation velocity (magenta solid line) in all the resolution cases.
As discussed in \citet{HSC21}, turbulence with a lower initial Mach number requires a higher Jeans resolution to capture the growth accurately; for example, $N_{\rm J} \geq 128$ is required for $\mathcal{M}_0=0.05$ in our previous results.
In this work, we find that $N_{\rm J}=32$ is enough resolution to accurately capture the growth of the turbulent velocity for $\mathcal{M}_0 = 0.5$.

\begin{figure}[htbp]
 \centering
  \plotone{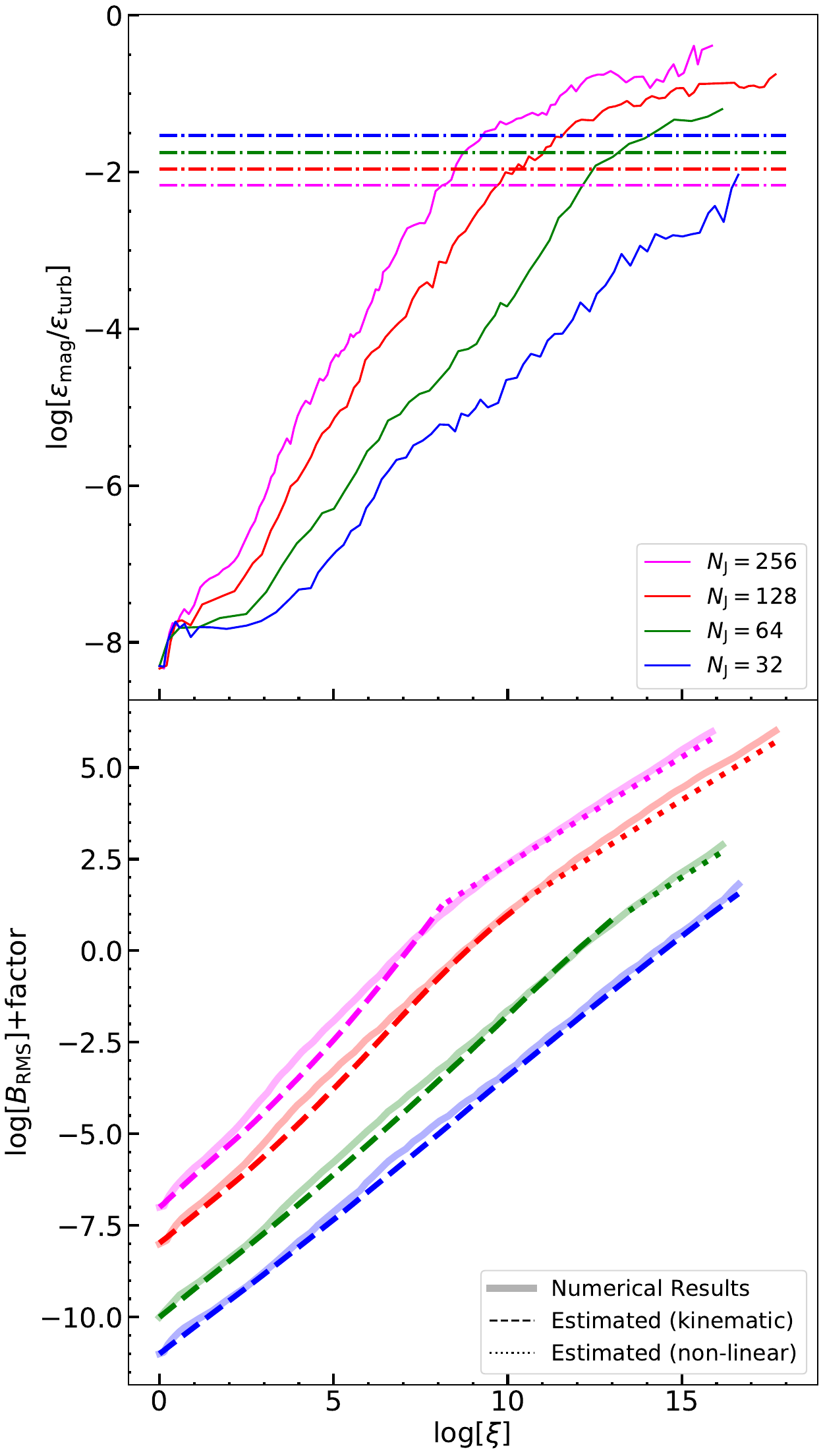}
  \caption{Upper panel: evolution of specific energy ratios between the magnetic energy and turbulent kinetic energy for different Jeans resolutions, as indicated in the legend. The dot-dashed lines show the estimated energy ratio at the onset of the non-linear stage.
  Lower panel: comparison between the simulation results and the analytic estimates in the kinematic (dashed) and non-linear (dotted) stages. We multiply factors to separate the different Jeans resolution cases, to facilitate the comparison between the simulations and the models.
  For $N_{\rm J}$ = 256, 128, 64, and 32, the factors are +2, +1, -1, and -2 in the logarithmic scale, respectively.
  }
    \label{fig:b_comparison_reso} 
\end{figure}

Figure~\ref{fig:b_comparison_reso} shows the evolution of the specific energy ratio (upper panel) and the comparison between the analytic estimates and our simulation results (lower panel) for various Jeans resolutions $N_{\rm J}$.

In the upper panel, the solid curves denote our simulation results, and the dash-dotted lines are the corresponding transition ratios obtained by Equation~(\ref{eq:R_nl0}), which are distributed between 0.0068 and 0.029.
We can see that the models for $N_{\rm J} \geq 64$ reach the non-linear stage, but not for $N_{\rm J}=32$.

In light of the above results, we present the comparison between our simulations and the estimated field strength obtained by Equations~(\ref{eq:B_kinematic1}) and~(\ref{eq:B_nonlinear}) in the lower panel.
The transition points in each color between the dashed and dotted curves correspond to the intersections between the solid curves and the dash-dotted lines in the upper panel. 
We add factors to the results for ease of viewing.
We can see that our simulation results can be well reproduced by our analytic estimates for various resolutions, both in the kinematic and non-linear stages.
The transition between the two stages is also consistent with the analytic estimate for all resolutions.
This fact indicates the robustness of our models.

\section{Discussion}\label{sec:discussion}

The magnetic field plays important roles also in the mass accretion phase after the collapse phase.
In a general scenario,
a collapsing primordial cloud becomes adiabatic when the density reaches $\gtrsim 10^{-5}$ g cm$^{-3}$, which is determined by the end of H$_2$ dissociation.
This threshold density is almost independent of the metallicity \citep{Omukai05}.
As a result, a hydrostatic core forms, known as a protostar, accompanied by an accretion disk \citep{Larson69,Penston69, Greif2012}.
This disk can fragment into pieces when it becomes gravitationally unstable due to the mass accretion from the surroundings, resulting in a multiple primordial stellar system \citep[e.g.,][]{Greif2012,Stacy16,Susa19}, similar to present-day star formation and fragmentation in disks \citep{Kuruwita19}.

Most of the recent numerical studies \citep{Sharda20, Sharda21_dynamo, Stacy22} have shown that strong magnetic fields amplified by the dynamo effect can moderately suppress the fragmentation of the disk, resulting in a relatively top-heavy mass function\footnote{Note that \cite{Hirano22} claimed the magnetic field amplification around the formed protostars may be so significant that fragmentation may be totally suppressed.}.

Our results indicate that strong magnetic fields always exist in PopIII prestellar cores irrespective of $\gamma_{\rm eff}$, i.e., different cooling/heating prescriptions or finite metallicity cases.
Thus, the theoretical argument that the expected IMF of the first stars would be more top-heavy than those in non-magnetized calculations becomes more robust than what is considered so far.

This fact might give some suggestions for the observational traces of the first stars.
If the dynamo-generated magnetic field indeed can suppress the disk fragmentation, the formed first stars would become more massive than expected by earlier numerical studies without magnetic effects.
This means that the formation of the low-mass first stars, which can survive until the present-day universe ($\lesssim 0.8 M_{\odot}$), tends not to be favored. 

On the other hand, the existence of the strong magnetic field at the onset of the accretion phase gives advantages to the formation of a massive tight population III star binary, which is considered to be a progenitor of the massive black hole binary that emits gravitational waves (GWs).
In order to coalesce to emit GWs within a Hubble time, the separation of the black hole binary has to be $\lesssim 100R_\odot$, which is as large as the expanded size of the progenitor protostars.
In the absence of strong magnetic fields, transportation of angular momentum from the spin of protostars to the orbital one results in a widening of the binary system\citep{Kirihara23}. Thus, it is hard to form such a close binary.
The magnetic breaking in the presence of a strong magnetic field could potentially counteract this widening mechanism by the spin-orbit angular momentum transfer.
So far, no studies have examined in detail the conditions for first stellar binary mergers incorporating the effects of magnetic braking. More in-depth research on this issue is needed.

\section{Summary}\label{sec:summary}
We have performed high-resolution numerical simulations to comprehensively study the magnetic field evolution associated with turbulence during the collapse of prestellar cores in the primordial star formation environment.
We have also generalized analytic estimates for the growth rate of the magnetic field during the gravitational contraction and compared them with our numerical simulation results.
Finally, we performed additional collapse simulations for the resolution study to reinforce our arguments.

Our main findings are summarized below.
\begin{enumerate}
    \item Amplification and saturation of the turbulent velocity through the gravitational contraction are not significantly affected by the back-reaction from the strong magnetic field (Figures~\ref{fig:velocity} and~\ref{fig:velocity_reso})

    \item The Kazantsev spectrum ($\propto k^{3/2}$) emerges in the collapsing clouds until the saturation of the magnetic field (Figure~\ref{fig:spectra}). 

    \item The magnetic/kinetic energy ratio at the saturation becomes a constant even if the turbulent velocity keeps increasing during the collapse, because the growth rate of the magnetic energy equals that of the turbulent kinetic energy at late times, and its value depends on $\gamma_{\rm eff}$. The smaller the $\gamma_{\rm eff}$, the larger the energy ratio, because the saturation turbulent velocity is smaller for lower $\gamma_{\rm eff}$ (Figure~\ref{fig:bratio}).

    \item The growth rates of the magnetic field due to the turbulent dynamo in both the kinematic stage and the non-linear stage depend on $\gamma_{\rm eff}$ (Equations~\ref{eq:B_kinematic1}, \ref{eq:depsilon_dt}, and Figure~\ref{fig:b-field}). 
    
    \item The transition specific energy ratio between the magnetic energy and the turbulent kinetic energy can be obtained solely from the numerical Jeans resolution (Equation~\ref{eq:R_nl0}), and it does not depend on $\gamma_{\rm eff}$.
   
    \item The magnetic field evolution in the collapsing gas clouds is well described by Equations~(\ref{eq:B_kinematic1}) and~(\ref{eq:B_nonlinear}) (c.f., Figures~\ref{fig:b_comparison} and~\ref{fig:b_comparison_reso}), for various $\gamma_{\rm eff}$ and numerical Jeans resolutions.

\end{enumerate}

Combined with the series of our previous studies \citep{HSC21,HSC22}, we confirm the growth of tangled magnetic fields as well as the supersonic turbulence in collapsing primordial clouds. This result is supported not only by high-resolution numerical simulations, but also by detailed analytic calculations. We find that the simulation and analytical results agree well with each other. Moreover, we extended the theory to general $\gamma_{\rm eff}$ cases, for which the agreement still holds at a very good level, providing support for the validity of the simulation and analytical results.

\software{\textsc{athena++} \citep{Athenapp,Tomida23}, \textsc{yt} \citep{yt}}

\acknowledgments
We are gratful to the anonymous referee for the careful reading of the manuscript and the constructive comments.
We thank Kengo Tomida, Kazuyuki Sugimura, Tsuyoshi Inoue, Kazutaka Kimura, Kohei Inayoshi, Sadanari Kenji Eric, Shingo Hirano, and Raiga Kashiwagi for fruitful discussions and useful comments.
This work was supported by JST SPRING, Grant Number JPMJSP2117.
We are thankful for the support by Ministry of Education, Science, Sports and Culture, Grants-in-Aid for Scientific Research No.22K03689.
C.~F.~acknowledges funding provided by the Australian Research Council (Future Fellowship FT180100495 and Discovery Project DP230102280), and the Australia-Germany Joint Research Cooperation Scheme (UA-DAAD). C.~F.~further acknowledges high-performance computing resources provided by the Leibniz Rechenzentrum and the Gauss Centre for Supercomputing (grants~pr32lo, pr48pi and GCS Large-scale project~10391), the Australian National Computational Infrastructure (grant~ek9) and the Pawsey Supercomputing Centre (project~pawsey0810) in the framework of the National Computational Merit Allocation Scheme and the ANU Merit Allocation Scheme.
Numerical calculations in this work were carried out on Yukawa-21 at the Yukawa Institute Computer Facility and Cray XC50 at Center for Computational Astrophysics, National Astronomical Observatory of Japan.
Computations and analysis described in this work were performed using the publicly-available \texttt{Athena++} and \texttt{yt} codes, which are the products of a collaborative effort of many independent scientists from numerous institutions around the world.


\bibliography{references}{}
\bibliographystyle{aasjournal}

\end{document}